\documentclass[journal]{IEEEtran}
\ifCLASSINFOpdf
\else
   \usepackage[dvips]{graphicx}
\fi
\usepackage{url}

\usepackage{setspace}
\usepackage{soul}
\usepackage{color, xcolor}
\usepackage{setspace}
\usepackage{graphicx}
\usepackage{subfig}
\usepackage{amssymb}
\usepackage{amsmath}

\usepackage{amsfonts}
\usepackage{graphicx}
\usepackage{epstopdf}
\usepackage{float}
\usepackage{stfloats}
\usepackage{indentfirst}
\usepackage[explicit]{titlesec}
\usepackage[citecolor=red]{hyperref}
\usepackage{caption}
\usepackage[ruled,linesnumbered]{algorithm2e}

\begin{document}
	
	\title{Intelligent MIMO Detection Using Meta Learning}
	\author{Haomiao Huo,
		    Jindan Xu,
		    Gege Su,
		    Wei Xu, \IEEEmembership{Senior Member, IEEE,}
		    and Ning Wang \IEEEmembership{Member, IEEE,}
	   \thanks{This work was supported in part by the National Key Research and Development Program 2020YFB1806600, and the NSFC under grants 62022026, 62211530108, and 61771431 (Corresponding author: Wei Xu).}
	   \thanks{H. Huo and W. Xu are with the National Mobile Communications Research Laboratory, Southeast University, Nanjing 210096, China (e-mail: 220190715@seu.edu.cn; wxu@seu.edu.cn).}
	   \thanks{J. Xu is with Engineering Product Development (EPD) Pillar, Singapore University of Technology and Design, Singapore 487372, Singapore, and she is also with the National Mobile Communications Research Lab, Southeast University, Nanjing 210096, China (email: jindan\_xu@sutd.edu.sg).}
       \thanks{G. Su and N. Wang are with the School of Information Engineering, Zhengzhou University, Zhengzhou 450001, China (e-mail: sgg@gs.zzu.edu.cn; ienwang@zzu.edu.cn).
       N. Wang is also with the State Key Laboratory of Millimeter Waves, Southeast University, Nanjing 210096, China.}}

\markboth{IEEE Wireless Communications Letters, Submitted}%
{Shell \MakeLowercase{\textit{et al.}}: A Sample Article Using IEEEtran.cls for IEEE Journals}

\maketitle
	
\begin{abstract}
 In a $K$-best detector for multiple-input-multiple-output (MIMO) systems, the value of $K$ needs to be sufficiently large to achieve near-maximum-likelihood (ML) performance.
 By treating $K$ as a variable that can be adjusted according to a fitting function of some learnable coefficients,
 an intelligent MIMO detection network based on deep neural networks (DNN) is proposed to reduce complexity of the detection algorithm with little performance degradation.
 In particular, the proposed intelligent detection algorithm uses meta learning to learn the coefficients of the fitting function for $K$
 to circumvent the problem of learning $K$ directly.
 The idea of network fusion is used to combine the learning results of the meta learning component networks.
 Simulation results show that the proposed scheme achieves near-ML detection performance while its complexity is close to that of linear detectors.
 Besides, it also exhibits strong ability of fast training.
\end{abstract}
	
	\setlength{\parskip}{0\baselineskip} 

\section{Introduction}
\IEEEPARstart{T}{he} past decades have witnessed tremendous successes in communication technologies, leading us to the 5G era.
The dependence on large antenna array and higher order modulation has increased the complexity of signal detection.
In multiple-input multiple-output (MIMO) systems, maximum-likelihood (ML) detector achieves the best error rate performance at the cost of high computational complexity,
which makes it difficult to implement in practice \cite{b1}.
Conversely, simple linear detection schemes, such as zero-forcing (ZF) and minimum mean squared error (MMSE),
usually suffer from a noticeable performance degradation compared to the optimal ML MIMO detector \cite{b2}.

Tree-search-based MIMO detectors have been proposed to effectively achieve near-ML detection performance \cite{b3}.
There are three main kinds of tree-search-based detection algorithms which are depth-first, width-first, and measurement-first algorithms.
The design of these algorithms aims at minimizing the search of leaf nodes without significant performance loss compared with ML detection.
The $K$-best algorithm is a typical width-first tree-search-based MIMO detection algorithm, where only $K$ nodes with the smallest measurements are retained at each layer.
In order to achieve near-ML detection performance, however, $K$ in the $K$-best algorithm has to be large.
It involves the sorting problem when retaining the $K$ nodes with the smallest measurements, which makes the implementation complex \cite{b4}.
	
Recently, deep learning (DL) has been successfully applied to various scenarios in communications.
It has been shown that DL can provide significant performance improvement for signal processing and communication problems,
such as decoding, demodulation, and detection \cite{b5} \cite{b6}.
In \cite{b8}, an end-to-end detector based on DNN was trained by treating the orthogonal frequency division multiplexing (OFDM) modulation and wireless channel as a black box.
The received signals, including a pilot block and a data block, are fed to the black-box DNN as input. The network outputs estimates of the transmitted signals.
Further in \cite{b9}, a DL method, namely Detection Network (DetNet), was proposed by following the calculation procedure of gradient-based optimization.
The DetNet is devised with low computational complexity and tries to approach the optimal performance.
A sphere decoding algorithm based on DL was studied in \cite{SD} where the radius of the decoding hypersphere is learned by a DNN prior to decoding.
In \cite{SVD}, the existing deep autoencoder for MIMO was improved by introducing singular-value decomposition (SVD) as a differentiable layer in the network.
However, the above DL-based methods may experience significant delay due to the need for online collection of sufficient training data \cite{b10}.
In addition, the long training time makes these methods prohibitive from adapting to time-varying communication scenarios.
	
Meta learning is an emerging learning technique which uses previous knowledge and experience to guide the learning process of new tasks \cite{b11}.
It aims at designing a network that can, through learning, quickly adapt to the new environment through a relatively small amount of new training data.
Motivated by these properties, an intelligent MIMO detection network using meta learning is proposed in this work.
Different from existing data-driven MIMO detection schemes, the meta learning network is used to learning a variable $K$ to achieve adaptive $K$-best MIMO detection
such that the computational complexity is reduced while little performance degradation is introduced.

Specifically, the value of $K$, instead of being a constant as in classical $K$-best algorithms, is treated as a variable at each layer of the search tree.
Such a scheme reduces the overall number of path nodes to be examined.
Moreover, to deal with the difficulty of learning $K$ directly, a fitting function of some learnable coefficients is designed to determine the value of $K$.
Meta learning is introduced to enable quick learning of the coefficients of the fitting function.
Network fusion is used to combine the learning results from multiple meta learning component networks
such that both detection performance and generalization ability can be improved, compared with a single meta learning model.
According to the $K$ values, a low-complexity DL network to select the path nodes intelligently and output estimates of the transmitted signals is designed.
	
\emph{Notations}: Throughout this letter, 
${( \cdot )}^{T}$ denotes transpose,
 ${( \cdot )}^{*}$ denotes matrix inversion,
$\left| \cdot \right|$ represents the absolute value operator, 
$\hat{\mathbf{x}}$ is an estimate of $\mathbf{x}$, and the Frobenius norm of vector $\mathbf{a}$ is denoted by $\left\| \mathbf{a} \right\|$.
	
	\section{System Model}
	\setlength\parindent{10pt}
	\setlength{\topskip}{0ex}
	\setlength{\parskip}{0ex}
	
	A spatial multiplexing MIMO system with $N_{t}$ transmit antennas and $N_{r}$ receiver antennas ($N_{r}\geq N_{t}$) is considered.
    The received baseband symbol vector is expressed as
	\begin{equation}
		\mathbf{y} = \mathbf{H}\mathbf{x} + \mathbf{n},
	\end{equation}
	where $\mathbf{x}=\left\lbrack x_{1},x_{2},\ldots,x_{N_{t}}\right\rbrack^{T}$ is the transmitted symbol vector,
    $\mathbf{y}=\left\lbrack y_{1},y_{2},\ldots,y_{N_{r}}\right\rbrack^{T}$ is the received symbol vector,
    $\mathbf{H}$ is the channel matrix from the transmitter array to the receiver array,
    and $\mathbf{n}=\left\lbrack n_{1},n_{2},\ldots ,n_{N_{r}}\right\rbrack^{T}$ is an additive noise term.
	
	For an ML detector, it selects the vector $\mathbf{x}$ that maximizes the channel transition probability, $P\left( \mathbf{y} \middle| \mathbf{x} \right)$,
    as the estimate of the transmitted symbol vector. The transition probability of the MIMO channel follows a multi-dimensional Gaussian distribution
	\begin{equation}\label{py_x}
	P\left( \mathbf{y} \middle| \mathbf{x} \right)
    = \frac{1}{\left( \pi N_{0} \right)^{N_{r}}}\exp\left( - \frac{1}{N_{0}}\left\| {\mathbf{y} - \mathbf{H}\mathbf{x}} \right\|^{2} \right),
	\end{equation}
	where $N_{0}\mathbf{I}$ is the covariance matrix of the zero-mean additive noise $\mathbf{n}$.
    According to (\ref{py_x}), maximizing $P\left( \mathbf{y} \middle| \mathbf{x} \right)$ is equivalent to
    minimizing $\left\| {\mathbf{y} - \mathbf{H}\mathbf{x}} \right\|^{2}$, which gives the form of ML detection
	\begin{equation}\label{py_y}
		\hat{\mathbf{x}} = {{\arg\min\limits_{\mathbf{x} \in \Omega}}\left\| {\mathbf{y} - \mathbf{H}\mathbf{x}} \right\|^{2}},
	\end{equation}
	where $\Omega$ represents the set of transmitted symbol vectors.
	
	In general, it is highly complex to directly solve for the ML detection results.
    To simplify the ML-based MIMO detection problem, QR decomposition of $\mathbf{H}$ is exploited \cite{b12}
	\begin{equation}
		\mathbf{H} = \mathbf{Q}\mathbf{R} = \mathbf{Q}\genfrac{[}{]}{0pt}{}{\bar{\mathbf{R}}_{N_{t} \times N_{t}}}{\mathbf{0}_{{\lbrack{N_{r} - N_{t}}\rbrack} \times N_{t}}} = \left\lbrack {\mathbf{Q}_{1}\mathbf{~}\mathbf{Q}_{2}} \right\rbrack\genfrac{[}{]}{0pt}{}{\bar{\mathbf{R}}}{\mathbf{0}}.
	\end{equation}

	The objective term in (\ref{py_y}) is thus further expressed as
	\begin{equation}\label{py_5}
		\begin{aligned}
			\left\| {\mathbf{y} - \mathbf{H}\mathbf{x}} \right\|^{2} = \left\| {{\mathbf{Q}^{*}_{1}}\mathbf{y} - \bar{\mathbf{R}}\mathbf{x}} \right\|^{2} + \left\| {{\mathbf{Q}^{*}_{2}}\mathbf{y}} \right\|^{2}.
		\end{aligned}
	\end{equation}
\noindent
    By defining $\mathbf{z} \triangleq {\mathbf{Q}^{*}_{1}}\mathbf{y}$ and substituting (\ref{py_5}) into (\ref{py_y}), the ML detection problem is rewritten as
	\begin{equation}\label{py_6}
		\begin{aligned}
			{\hat{\mathbf{x}}} = {\arg\min\limits_{\mathbf{x} \in \Omega}}{\sum\limits_{i = 1}^{N_{t}}\left( {z_{i} - {\sum\limits_{j = 1}^{N_{t}}{r_{ij}x_{j}}}} \right)}^{2},
		\end{aligned}
	\end{equation}
    where $z_{i}$ is the $i$-th element of the vector $\mathbf{z}$, $x_{j}$ is the $j$-th element of the vector $\mathbf{x}$,
    and $r_{ij}$ is the ($i,j$)-th entry of the matrix $\bar{\mathbf{R}}$.

     According to (\ref{py_6}), (\ref{py_y}) is transformed into a minimum path search problem for a weighted tree.
     The $j$-th element of the transmitted symbol vector $\mathbf{x}$ corresponds to the $j$-th layer of the tree.
     Each node of the $j$-th layer is uniquely determined by the path from the root to this node, which has a path metric
	\begin{equation}\label{py_7}
		\delta\left( x_{j} \right) = \delta\left( x_{j + 1} \right) + \sigma\left( x_{j} \right) ,
	\end{equation}
    where $\sigma\left( x_{j} \right) \triangleq \| {z_{j} - {\Sigma_{i = j}^{N_{t}}{r_{ji}x_{i}}}} \|^{2}$ is the node metric.
    Note that the root node corresponds to $j = N_{t}$, and $\sigma\left( x_{{N_{t}}} \right) = 0$.
	According to (\ref{py_7}), the path metric is non-decreasing from the root to the leaf along any path.
    Therefore, ML detection is equivalent to finding the leaf with the minimum path metric in the search tree.
	
	The $K$-best algorithm is a representative of the width-first tree-search-based MIMO detection scheme.
    Its basic idea is that only $K$ nodes with the smallest measurements are retained and examined at each layer.
    The steps of the $K$-best algorithm are outlined in Algorithm \ref{algorithm}.
	\begin{algorithm}[t]
		\caption{$K$-best algorithm}\label{algorithm}
		\KwIn{matrix $\mathbf{H}$, vector $\mathbf{z}$}
		initialization: $\delta\left( x_{j} \right) = 0, \sigma\left( x_{j} \right)= 0, j = N_{t}$\;
		\While{$j\geq 1$}
		{$K$ paths saved in layer $j$ is corresponding to $K$ sub vectors: $x_{(j)}^{1}, x_{(j)}^{2},\ldots, x_{(j)}^{K}$\;
			expand the $K$ sub vector to $K \times Q$ child nodes in layer $j-1$, $Q$ denotes the modulation order\;
			calculate the path metrics in layer $j-1$ $\delta\left( x_{j-1} \right) = \delta\left( x_{j} \right) + \sigma\left( x_{j-1} \right)$\;
		    save $K$ nodes with minimum path metrics, delete other child nodes\;
			$j=j-1$\;}
		Take the path with the smallest path metric as the final detection result\;
		\KwOut{estimated transmitted vector $\hat{\mathbf{x}}$}
	\end{algorithm}

\vskip12pt

	\section{MIMO Detection Network Design}
	
     In this section, we propose an intelligent detection network using meta learning to achieve adaptive $K$-best-based MIMO detection.
     Since fewer nodes need to be retained and examined at layers closer to the root,
     it is natural to design a mechanism with varying $K$ for different layers in the search tree
     such that $K$ is minimized at each layer to reduce complexity while still achieving near-ML detection performance.
     In general, learning the minimum value of $K$ directly is challenging, even through a straightforward data-driven DL approach.
     To circumvent this issue, we propose to design a parametric learnable fitting function to obtain the value of $K$.
     A meta learning approach is introduced to learn the coefficients of this fitting function.
     Convolutional neural networks (CNNs) are then constructed to conduct the subsequent path node selection procedure.

     The proposed intelligent MIMO detection network therefore contains two modules, the coefficient training module for $K$ determination
     and the path selection module for signal detection.
     The output of the coefficient training module based on meta learning is fed to the parametric fitting model to obtain an integer $K$ for each layer,
     which is used for the construction of the CNNs in the DL-based path selection module.
     The inputs of the CNNs, which include the received signals and the channel state information, are obtained through pre-processing.
     The convolutional networks output the saved path nodes to the path metric calculation unit.
     The DL path selection module finally calculates and selects the estimates of the transmitted signal.

		\begin{figure}[t]
		\begin{center}
			\includegraphics[scale=0.3]{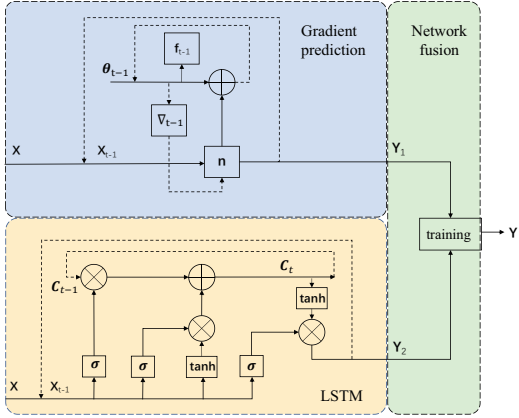}
		\end{center}
		\caption{Structure of the coefficient training module using meta learning.}
		\label{fig:1}
    \vskip-6pt
	\end{figure}

\subsection{Meta-Learning-Based $K$ Learning and Network Fusion}
	The complexity of a $K$-best algorithm is primarily determined by the value of $K$.
    In order to achieve a near-ML detection performance, a large $K$ value should be used, which results in high computational complexity.
    While a smaller value of $K$ generally results in degraded detection performance.

    In the $K$-best algorithm, it has been verified that the minimum $K$ corresponding to the optimal search path
    grows with the layer index number (away from the root).
    Specifically, in a layer close to the root node, a smaller $K$ can be used without deviating from the optimal path,
    while in a layer far from the root, a larger $K$ should be used instead.
    Therefore, $K$ can be a design variable that takes different values at different layers of the search tree.
    In this way, the optimal search path can be attained while fewer nodes are examined by the algorithm.
    The proposed modified $K$-best detection scheme thus achieves near optimal performance with reduced complexity.
    Considering that it is difficult to obtain the optimal $K$ value directly, we propose to learn the value of $K$ in each layer by training a DNN
    to approximate a fitting function for $K$ with learnable coefficients $a$, $b$ and $c$.
    The fitting function has the form
	\begin{equation}\label{py_8}
		K = a \times k^{b} + c,
	\end{equation}
	where $k$ denotes the layer index number away from the root.
    Once the network is trained, the value of $K$ is selected as the integer closest to the result given by (\ref{py_8}) for each layer.
	
	The meta learning method is introduced to learn the coefficients $a$, $b$ and $c$ in (\ref{py_8}) for predicting the minimum $K$.
	The purpose of meta learning is to train a function that helps train the task-specific parameters in the model.
	Meta learning enables the model to learn the network parameter adjustments based on hyper-parameters\cite{b11}.
    As a result, it can quickly adapt to new environment with the assistance of existing knowledge.

	We design a coefficient learning module using meta learning to train the coefficients of the fitting function (\ref{py_8}).
    The initial value of the learnable coefficients $a$, $b$ and $c$ are defined as the input $\mathbf{X}$ of the network, i.e.
	\begin{equation}
		\mathbf{X} = \left( {a,b,c} \right).
	\end{equation}
\noindent
	Fig. \ref{fig:1} shows the structure of the meta learning-based coefficient training module.
    Both meta learning component networks, including the predictive gradient network and the long short-term memory (LSTM) network,
    take $\mathbf{X}$ as the input.
    After initialization and training, the two networks output updated coefficients $\mathbf{Y}_{1}$ and $\mathbf{Y}_{2}$, respectively.
    Network fusion is used to combine the component outputs $\mathbf{Y}_{1}$ and $\mathbf{Y}_{2}$ to give the final output $\mathbf{Y}$ of the coefficient training module.

		\begin{figure}[t]
		\begin{center}
			\includegraphics[scale=0.35]{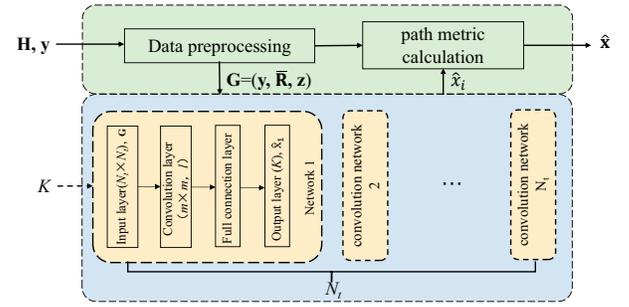}
		\end{center}
		\caption{Structure of the DL path selection module.}
		\label{fig:2}
    \vskip-6pt
	\end{figure}

	Specifically, the upper meta learning component network as shown in Fig. \ref{fig:1} is gradient prediction.
    $\mathbf{\theta}$ is the meta learning network state information, and $\nabla$ is the prediction of gradient.
    Through gradient prediction, a faster and more accurate neural network optimizer is obtained.
    The input $\mathbf{X}$ is updated to the output $\mathbf{Y}_{1}$ through the neural network optimizer.
    The lower meta learning component network is an LSTM network, which is a special cyclic neural network.
    It deletes or adds information to cell state through a gate structure.
    The learnable coefficients in $\mathbf{X}$ are used as inputs of the LSTM network to obtain the coefficient update result $\mathbf{Y}_{2}$.
    Both meta learning component networks use hyper-parameters to guide network training and use their outputs as a new input for the next round of training,
    which enables faster training. 
	We denote the two component network structures as
	\begin{equation}
		\mathbf{Y}_{1} = F_{1}\left( \mathbf{X} \right),
	\end{equation}
	\begin{equation}
		\mathbf{Y}_{2} = F_{2}\left( \mathbf{X} \right),
	\end{equation}
	where $F_{1}$ and $F_{2}$ are the network functions of the two meta learning component networks.

	The network fusion module in Fig. \ref{fig:1} is designed to combine the learning results of multiple meta learning component networks with different structures 
    to improve the detection performance and the ability of generalization. 
    The network fusion function is expressed as
	\begin{equation}
		\mathbf{Y} = {\omega_{1}\mathbf{Y}}_{1} + {\omega_{2}\mathbf{Y}}_{2},
	\end{equation}
	where $\omega_{1}$ and $\omega_{1}$ are two auxiliary learnable weighting scalars associated with the two component networks.
	
	As the input of the coefficient training module, learnable coefficient $\mathbf{X}$ is updated to $\mathbf{Y}_{1}$ and $\mathbf{Y}_{2}$
    through the two meta learning component networks.
    Then the outputs of the two component networks are combined by the network fusion module.
    The network fusion module trains the auxiliary learnable weighting scalars $\omega_{1}$ and $\omega_{1}$ to optimize the output $\mathbf{Y}$.

	\subsection{DL-Based Path Selection}
	
	In the previous section, appropriate value of $K$ is obtained by substituting the output $\mathbf{Y}$ of the coefficient training module into (\ref{py_8}).
    In this part, design of the intelligent search path selection module is investigated based on path metric calculation and CNNs.
		
    Fig. \ref{fig:2} shows the structure of the proposed DL path selection module.
    The path metric calculation unit first pre-processes the input data, which applies $\mathbf{Q}\mathbf{R}$ decomposition to the channel matrix $\mathbf{H}$
    and obtains $\mathbf{H} = \mathbf{Q}\mathbf{R}$.
    The vector $\mathbf{z} \triangleq {\mathbf{Q}^{*}_{1}}\mathbf{y}$ is calculated according to the decomposed matrix and the received signal vector.
    The received vector $\mathbf{y}$, the vector $\mathbf{z}$ and the matrix $\bar{\mathbf{R}}$ containing path measurement information are fed back to the convolutional networks.

	We design $N_{t}$ CNNs to replace the optimal path selection at each layer of the $K$-best MIMO detection algorithm.
    The number of CNNs is consistent with the dimension of the transmitted vector $\mathbf{x}$.
    As shown in Fig. \ref{fig:2}, $\mathbf{G} = \left( \mathbf{y}, \bar{\mathbf{R}}, \mathbf{z} \right)$ are the inputs of every convolutional network.
    The input layer is followed by $l$ convolutional layers and the size of the convolutional kernel is $m \times m$.
    The dimension of the next full connection layer is $m \times n$ and the ReLU function is used as the activation function.
    The dimension of the output layer is $K$, which corresponds to the number of path nodes to be examined in this layer of the search tree.
    $N_{t}, m, n, l$ and $K$ are positive integers and each parameter can be adjusted according to the use requirements.
    The convolutional neural network is expressed as
	  \begin{equation}
	 	{\hat{x}}_{i} = \Phi_{i}\left( \mathbf{G},\lambda \right),
	 \end{equation}
	 where $\Phi$ denotes the convolutional network function and $\lambda$ is the vector containing all network parameters.

	The output of every convolutional network is fed to the path metric calculation unit,
    which calculates the path metrics according to (\ref{py_7}).
    Finally, the DL path selection module outputs the estimated detection result $\hat{\mathbf{x}}$, which achieves the smallest path metric among all the search paths.

	\begin{figure*}[tbp]
		\begin{center}
			\includegraphics[scale=0.45]{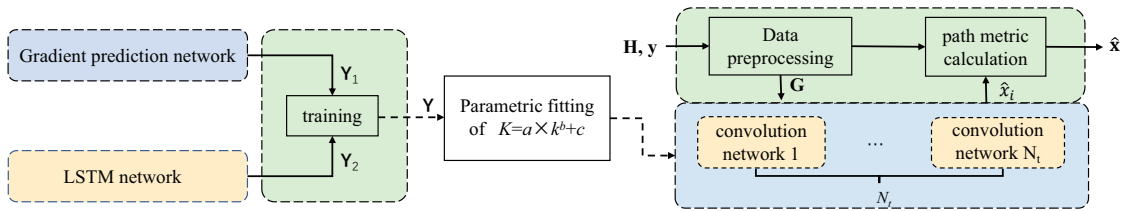}
		\end{center}
		\caption{Structure of the proposed intelligent MIMO detection network using meta learning.}
		\label{fig:3}
    \vskip-6pt
	\end{figure*}
	
    \subsection{Network Training}	
	The proposed intelligent MIMO detection meta network consists of three parts. 
    As illustrated in Fig. \ref{fig:3}, they are the coefficient training module, 
    the parametric fitting model and the DL path selection module.

	The network is used as a MIMO detector. 
    By inputting the received signal $\mathbf{y}$ and the channel matrix $\mathbf{H}$, the network outputs estimate $\hat{\mathbf{x}}$ of the transmitted signal vector $\mathbf{x}$.
    In the training phase, the coefficient training module is independently trained according to the channel matrix $\mathbf{H}$.
    The input to the network is an array that stores the learnable coefficients.
    The minimum number of reserved nodes per layer determined according to channel information is used to continuously optimize the learnable coefficient of the function.
    Then the coefficient training module and the DL path selection module are jointly trained,
    aiming at minimizing the gap between the detection result $\hat{\mathbf{x}}$ and the signal $\mathbf{x}$ actually transmitted.
    The transmitted and received data of each frame is a binary sequence related to the frame length.
    The
    training data is generated by channel simulation and the generated data is divided into a training set, a verification
    set and a test set. 
    The loss function is designed as
	\begin{equation}\label{py_14}
		{\rm Loss}\left( \lambda \right) = \frac{1}{N_{t}}{\sum\limits_{i = 1}^{N_{t}}\left\| {x_{i}-\Phi_{i}\left(\mathbf{G},\lambda \right)} \right\|^{2}}.
	\end{equation}
    The learning rate of the network adopts a gradual value in order to approach the global optimal solution.
    The model parameters of network are saved at the end of the training phase for implementation.
	

\section{Simulation Results}
	
		\begin{figure*}[htbp]
	\subfloat[16QAM]{
		\label{1}
		\begin{minipage}[c]{.5\linewidth}
			\centering
			\includegraphics[width=0.85\textwidth]{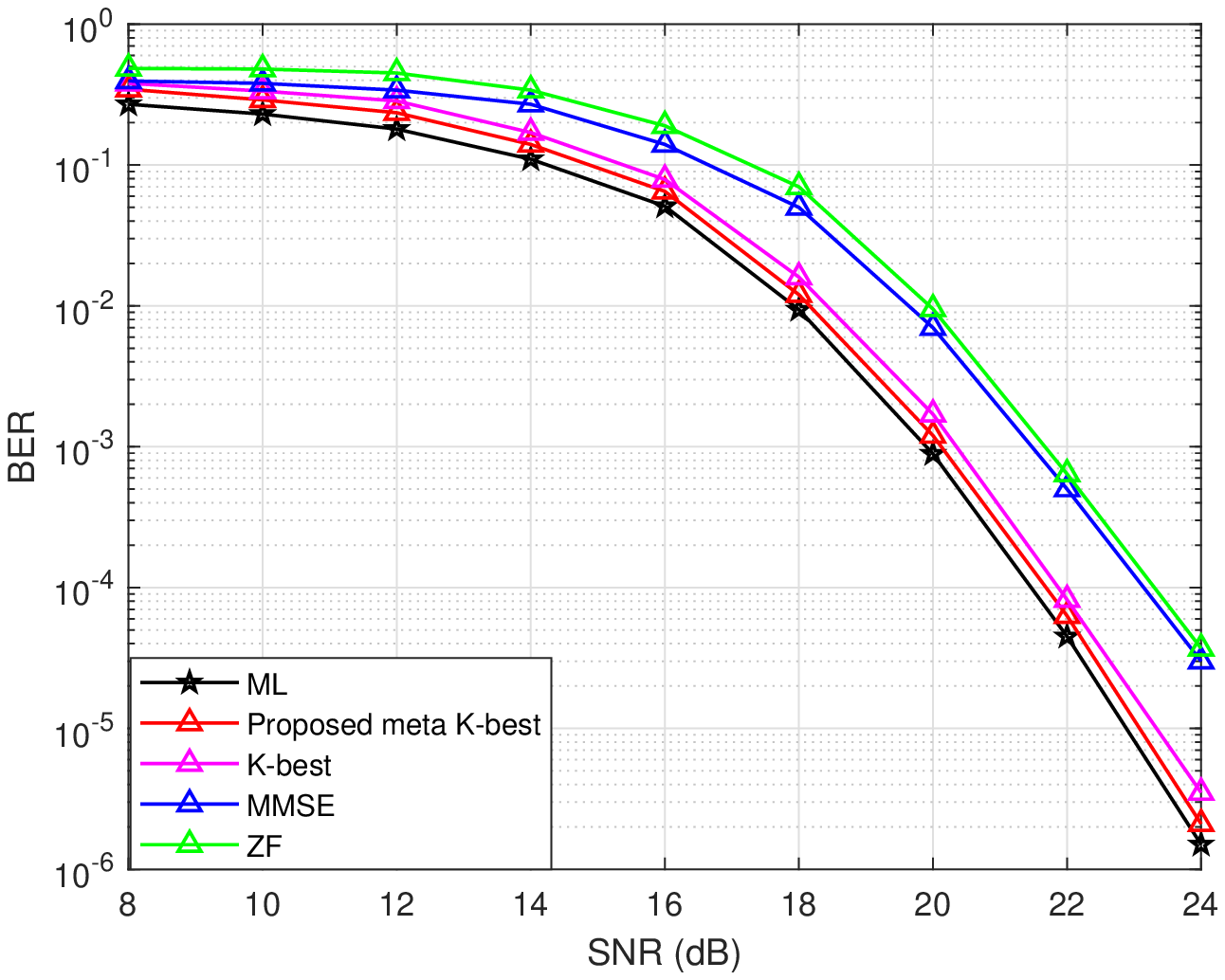}
		\end{minipage}
	}
	\subfloat[64QAM]{
		\label{2}
		\begin{minipage}[c]{.5\linewidth}
			\centering
			\includegraphics[width=0.85\textwidth]{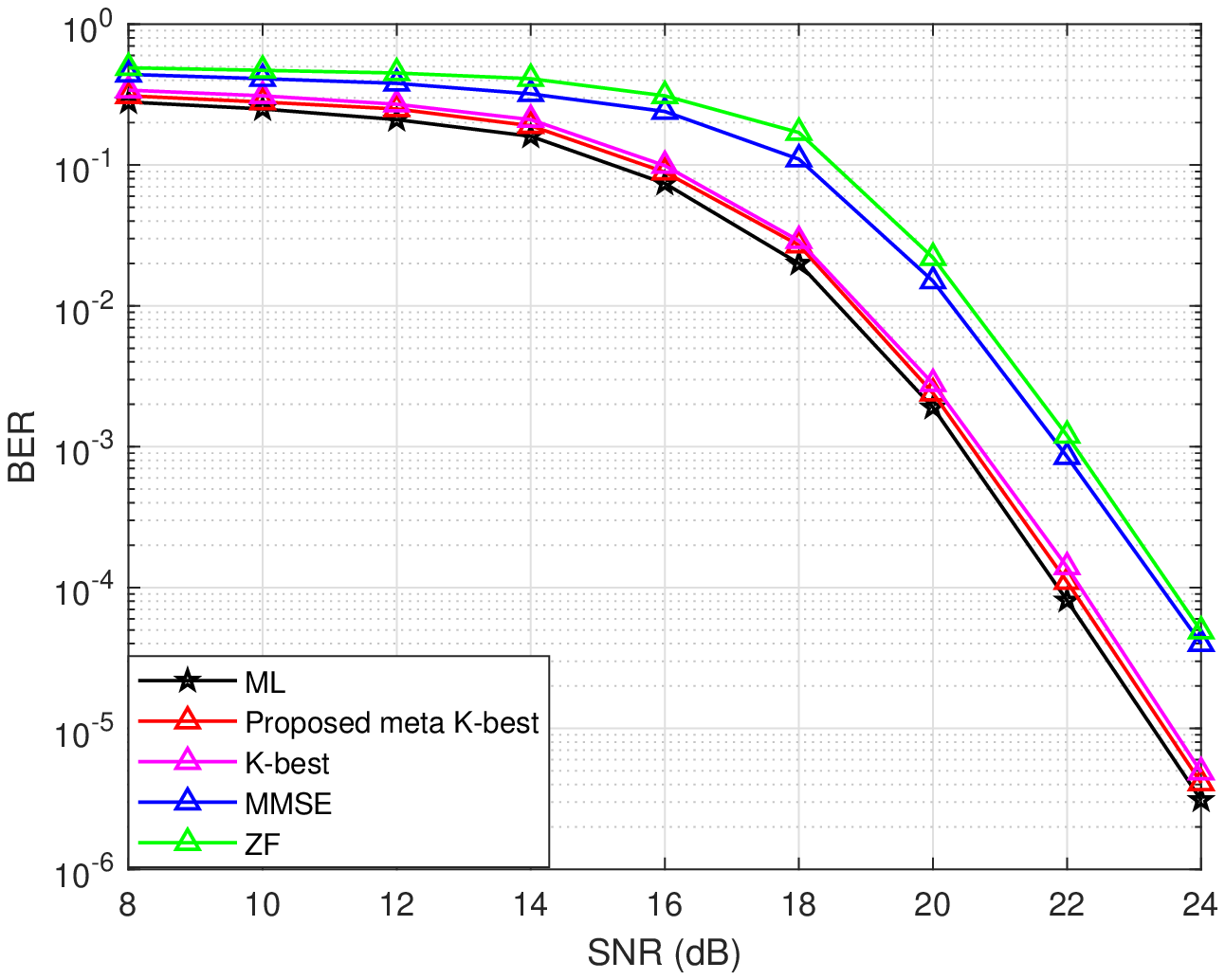}
		\end{minipage}
	}
	\caption{Performance comparison of the proposed MIMO detection network, ML, $K$-best, MMSE and ZF algorithms.}
	\label{fig:4}
\vskip-6pt
\end{figure*}

\begin{figure}[tp]
	\begin{center}
		\includegraphics[scale=0.53]{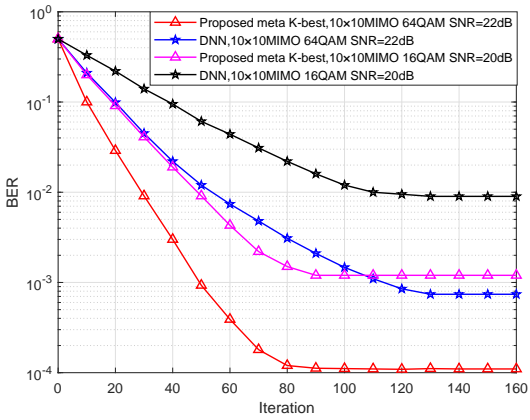}
	\end{center}
	\caption{Convergence of training of the proposed MIMO detection network and conventional DNN.}
	\label{fig:5}
\vskip-6pt
\end{figure}

	In this section, we evaluate the performance of the proposed intelligent MIMO detection network
    based on the $K$-best algorithm and meta learning through simulation experiments.
	
	Consider a 10$\times$10 spatial multiplexing MIMO system where 16-QAM and 64-QAM are employed. Gray code mapping is adopted for transmitted symbol.
    The additive white noise is modeled as a complex-valued Gaussian random variable with zero-mean for each receiver antenna.
    The network adopts 4 convolutional layers with the convolutional kernel size of $3 \times 3$.
    In the training phase, 1000 batches of training data are used.
    The learning rate, $\eta$, of the Adam optimization algorithm is set as the change learning rate, and the initial value is 0.01.
	
	Fig. \ref{fig:4} shows the bit error ratio (BER) of the proposed intelligent MIMO detection network with respect to the average signa-to-noise-ratio (SNR).
    The performances of the ML, MMSE, ZF and $K$-best schemes are also shown for comparison.
    The proposed MIMO detection network exhibits a BER performance close to that of ML detection over a wide range of SNR,
    which significantly outperforms the $K$-best and MMSE detection algorithms.
    When 16-QAM is employed, the proposed MIMO detection network achieves 0.39 dB, 1.81 dB and 2.07 dB average performance gain compared with $K$-best detection, MMSE detection
    and ZF detection respectively, while its performance is 0.29dB lower than ML detection.
    When 64-QAM is employed, the average performance gain is 0.26 dB, 1.93 dB and 2.14 dB advantage over the $K$-best, MMSE and ZF detection, respectively.
    The performance gap between the proposed scheme and ML detection is 0.37dB.
	
	In Fig. \ref{fig:5}, convergence of training process of the proposed MIMO detection network is compared with the conventional DNN.
    It shows that the proposed intelligent MIMO detection network converges much faster than the DNN over a wide range of SNR.
    The proposed MIMO detection network can therefore better adapt to fast changes to the communication scenario.
	
	Table \ref{x} compares the computational complexity of the proposed intelligent MIMO detection network with that of the classical detection algorithms\cite{b13},
    where $N_{t}$ denotes the number of transmit antennas, $Q$ denotes the modulation order, and $K$ is the number of the saved path nodes in each layer.
    As can be observed, the complexity of the proposed intelligent MIMO detection network is on the same order as that of linear detection algorithms
    such as MMSE and is significantly lower than that of ML detection and the conventional $K$-best detection.

\vskip12pt
	
    \section{Conclusion}
	A low-complexity intelligent MIMO detection network based on meta learning has been proposed.
    To reduce the computational complexity of the tree-search-based $K$-best MIMO detection algorithm, the proposed network treats $K$ as a variable
    and designs a fitting function to obtain an appropriate value of $K$ for each layer of the search tree.
    A coefficient training module applies meta learning and network fusion to learn the coefficients of the fitting function for $K$.
    The DL path selection module has been designed to intelligently select path nodes of the search tree to achieve near-ML detection performance.
    It has been shown through simulations that the error rate performance of the proposed intelligent MIMO detection network is close to that of ML detection
    for high-dimensional problems with low computational complexity and fast convergence properties.

	\begin{table}[t]
	\caption{Complexity of the proposed MIMO detection network}
	\label{x}
	\centering
	\small
	\begin{tabular}{l|l}
		\hline
		Detection algorithms & Complexity\\
		\hline
		ML & $O\left( Q^{N_{t}} \right)$\\
		MMSE & $O\left( N_{t}^{3} \right)$ \\
		$K$-best & $O\left( K2^{N_{t}} \right)$ \\
		Proposed meta $K$-best detection & $O\left( KN_{t}^{3} \right)$ \\
		\hline
	\end{tabular}
\end{table}

\end{document}